# Hidden phase uncovered by ultrafast carrier dynamics in thin Bi$_2$O$_2$Se

*Hao Li, Adeela Nairan, Xiaoran Niu, Yuxiang Chen, Huarui Sun, Linqing Lai, Jingkai Qin, Leyang Dang, Guigen Wang, Usman Khan*, Feng He**

H. Li, X. Niu, F. He

State Key Laboratory on Tunable Laser Technology, Harbin Institute of Technology (Shenzhen), Shenzhen 518055, P. R. China

Guangdong Provincial Key Laboratory of Aerospace Communication and Networking Technology, Harbin Institute of Technology (Shenzhen), Shenzhen 518055, P. R. China

E-mail: hefeng2020@hit.edu.cn

A. Nairan, U. Khan

Institute of Functional Porous Materials, School of Materials Science and Engineering, Zhejiang Sci-Tech University, Hangzhou 310018, P. R. China

E-mail: usmankhan.2017@tsinghua.org.cn

Y. Chen, H. Sun

School of Science and Ministry of Industry and Information Technology, Key Laboratory of Micro-Nano Optoelectronic Information System, Harbin Institute of Technology (Shenzhen), Shenzhen, Guangdong 518055, P. R. China

L. Lai, J. Qin

Sauvage Laboratory for Smart Materials, School of Materials Science and Engineering, Harbin Institute of Technology (Shenzhen), Shenzhen 518055, P. R. China

L. Dang, G.Wang

Shenzhen Key Laboratory for Advanced Materials, School of Materials Science and Engineering, Harbin Institute of Technology (Shenzhen), Shenzhen 518055, P. R. China

Keywords: ferroelectric transition, Bi$_2$O$_2$Se thin films, pump-probe experiment

Bi$_2$O$_2$Se has attracted intensive attention due to its potential in electronics, optoelectronics, as well as ferroelectric applications. Despite that, there have only been a handful of experimental studies based on ultrafast spectroscopy to elucidate the carrier dynamics in Bi$_2$O$_2$Se thin films, Different groups have reported various ultrafast timescales and associated mechanisms across films of different thicknesses. A comprehensive understanding in relation to thickness and fluence is still lacking. In this work, we have systematically explored the thickness-dependent Raman spectroscopy and ultrafast carrier dynamics in chemical vapor deposition (CVD)-grown Bi$_2$O$_2$Se thin films on mica substrate with thicknesses varying from 22.44 nm down to 4.62 nm at both low and high pump fluence regions. Combining the thickness dependence and fluence dependence of the slow decay time, we demonstrate a ferroelectric transition in the thinner (< 8 nm) Bi$_2$O$_2$Se films, influenced by substrate-induced compressive strain and non-equilibrium states. Moreover, this transition can be manifested under highly non-equilibrium states. Our results deepen the understanding of the interplay between the ferroelectric phase and semiconducting characteristics of Bi$_2$O$_2$Se thin films, providing a new route to manipulate the ferroelectric transition.

## 1. Introduction

In contrast to van der Waals (vdW) forces, Bi$_2$O$_2$Se exhibits relatively strong electrostatic forces between layers due to its alternately stacked positively charged [Bi$_2$O$_2$]$^{2+}$ and negatively charged Se$^{2-}$ layers along the c direction.[1, 2] With a moderate bandgap ranging from 0.8 eV to 2.09 eV,[3, 4] high mobility of 28900 cm$^2$ V$^{-1}$ s$^{-1}$ at 1.9 K and 450 cm$^2$ V$^{-1}$ s$^{-1}$ at room temperature,[5, 6] along with high thermal and chemical stability,[4, 7] Bi$_2$O$_2$Se holds great promise as a candidate for next-generation electronics,[8, 9] optoelectronics,[7, 10, 11] thermoelectrics,[12, 13] as well as ferroelectrics.[14, 15] Furthermore, developing controllable large-area and high-quality 2D Bi$_2$O$_2$Se films speeds up the applications.[7, 10]

Given the high demand for data storage, extensive research has focused on ferroelectric properties in Bi$_2$O$_2$Se.[16-18] Ghosh et al. reported the presence of ferroelectricity in 2 nm single-crystalline Bi$_2$O$_2$Se nanosheets through dielectric measurements and piezoresponse force spectroscopy.[16] This study highlights the strong correlation between the ferroelectric phase transition in Bi$_2$O$_2$Se and its atomic displacement. Subsequently, Zhu et al. employed ab initio calculations of the phonon-limited intrinsic mobility of Bi$_2$O$_2$Se, revealing a connection between its high mobility and an emerging interlayer ferroelectric phase transition.[19] They proposed that a low-frequency transverse optical (TO) mode with 6.8 meV (1.6 THz, IR active) significantly contributes to the large dielectric constant, and a slight biaxial elastic strain of less than 1.7% could soften this TO mode, triggering the ferroelectric transition, enhancing the dielectric constant, and consequently increasing mobility. Meanwhile, Wang et al. utilized coherent phonon spectroscopy to observe a symmetry-forbidden $A_{1g}$ mode (1.5 THz) and identified the formation of ferroelectric polarons in a 7 nm-thick $Bi_2O_2Se$ film.[20] Therefore, it is intriguing to investigate the ferroelectric transition in relation to the thickness-dependent ultrafast carrier dynamics.

Despite the limited number of experimental studies utilizing ultrafast spectroscopy to investigate the carrier dynamics in $Bi_2O_2Se$ thin films, various groups have reported a diversity of ultrafast timescales and their associated mechanisms. Tian et al. associated the rapid recovery time (several picoseconds) with electron-electron coupling in solution-processed $Bi_2O_2Se$ with a thickness of 155.5 nm.[1] Yu et al. attributed the fast recovery component (~1.46 ps) to excimer formation, 8.24 ps to hot excimer cooling, and the slow decay process (~122 ps) to excimer dissociation in the few-layer CVD-grown $Bi_2O_2Se$/mica nanosheets with a thickness of 11.3 nm.[21] Liu et al. linked the fast decay (~1 ps) to carrier defect trapping, and the slower decay (~100 ps) to defect-assisted Auger recombination in CVD-grown $Bi_2O_2Se$ nanoplates with a thickness of 10 nm.[22] Han et al. attributed both the fast (~8 ps) and slow (~90 ps) lifetimes to the phonon-assisted carrier relaxation processes in a CVD-grown 27.5 nm-thick sample.[23] With various timescales observed in different thicknesses, direct comparisons become challenging. It is, therefore, imperative to systematically investigate thickness-dependent ultrafast carrier dynamics. Additionally, comprehending high-density carrier behaviors remains crucial for applications in high-power electronics.

In this work, we systematically investigated the thickness-dependent and pump fluence-dependent ultrafast dynamics of thin $Bi_2O_2Se$ flakes on mica substrate prepared by the CVD method.[24] Employing degenerate pump-probe spectroscopy at a wavelength of 800 nm and Raman spectroscopy, we revealed a ferroelectric transition in the thinner (< 8 nm) $Bi_2O_2Se$ films under compressive strain from the substrate. This transition prominently manifests under conditions of high non-equilibrium. Our findings offer valuable insights for the understanding and design of optoelectronic and memory devices utilizing $Bi_2O_2Se$ thin films.[14]

## 2. Results and Discussion

## 2.1. Characterization of CVD-grown Bi$_2$O$_2$Se Thin Films

In a Bi$_2$O$_2$Se crystal, the atoms are arranged in an orthogonal lattice (a=b=3.88 Å, c=6.08 Å), as illustrated in Figure 1a. The tetragonal Bi$_2$O$_2$Se structure is alternately stacked by positively charged [Bi$_2$O$_2$]$^{2+}$ and negatively charged Se$^{2-}$ layers along the c direction, exhibiting inversion symmetry. Figure 1b shows the CVD-grown Bi$_2$O$_2$Se flakes with square or rectangular shapes.[24]

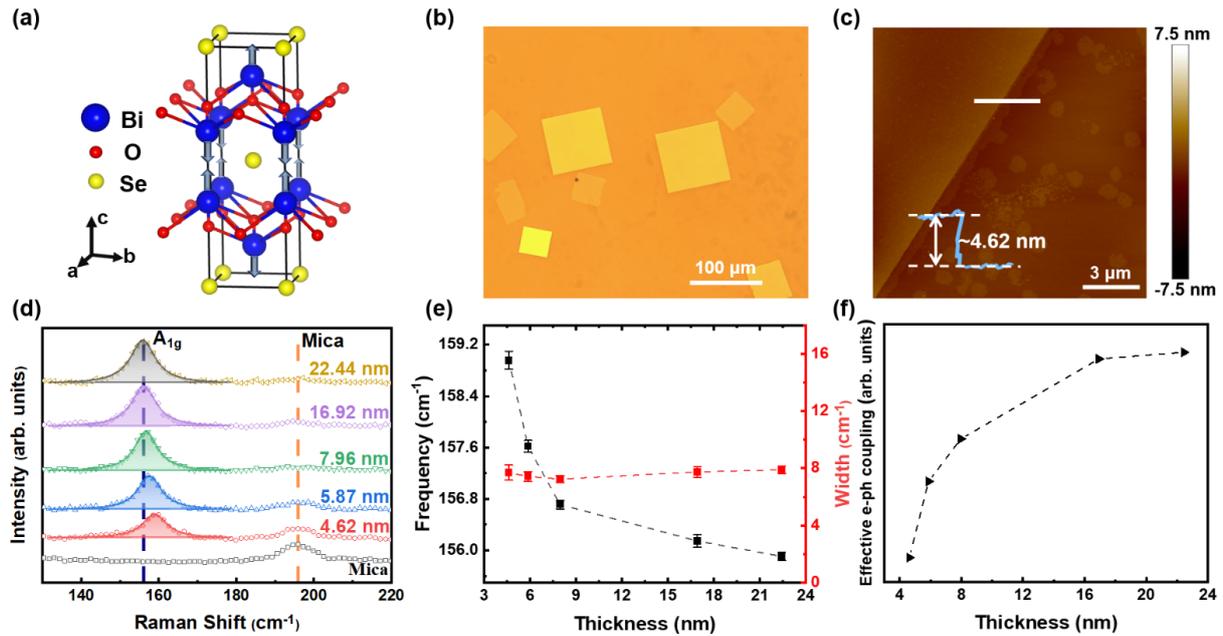

**Figure 1.** Characterization of Bi$_2$O$_2$Se thin films on mica. a) Crystal structure of Bi$_2$O$_2$Se thin films synthesized by the CVD method and atomic displacements of $A_{1g}$ vibrational modes, with the Bi, O, and Se atoms represented by blue, red, and yellow spheres, respectively. b) Optical microscopy image of the Bi$_2$O$_2$Se thin films. c) Atomic force microscope (AFM) topography of a multilayer Bi$_2$O$_2$Se with a thickness of 4.62 nm. d) Layer-dependent Raman spectra (empty dots) of Bi$_2$O$_2$Se thin films and the corresponding Lorentz fitting analysis (solid lines). Different colors represent different thicknesses, and dashed lines are included to guide the eye, representing the $A_{1g}$ mode in Bi$_2$O$_2$Se and the signature peak in mica. e) The fitted frequency (black squares) and the width (red squares) for $A_{1g}$ modes as a function of film thickness. f) The relationship between effective electron-phonon (e-ph) coupling and thickness.

The AFM measurement reveals that the thickness of a representative Bi$_2$O$_2$Se thin film is approximately 4.62 nm (~7 layers with monolayer thickness to be ~0.61 nm),[5] with the root-mean-square surface roughness to be 3.5 Å. All other thin films measured in this work are present in Figure S1 (Supporting Information). We have measured the Raman spectra in five different films on mica substrate, with thickness varying from 22.44 nm to 4.62 nm. The corresponding results are presented in Figure 1d. The characteristic peaks observed at 159 cm$^{-1}$ to 156 cm$^{-1}$ can be attributed to the $A_{1g}$ vibrational modes of Bi$_2$O$_2$Se flakes,[25] while the peak at 195.9 cm$^{-1}$ in the thinner films originates from the mica substrate, as indicated by the dashed lines.[7] It is worth noting that the mica peak is only visible in the thinner films. Through Lorentz fitting, we obtained the $A_{1g}$ phonon frequency and width as a function of film thickness, as depicted in Figure 1e. The $A_{1g}$ phonon frequency exhibits a blue-shift in thinner films, while the width remains relatively unaffected by film thickness. This frequency trend is consistent with previous reports,[7, 13, 24] suggesting that it may be induced by an out-of-plane compressive strain at the Bi$_2$O$_2$Se /Mica interface.[26] Moreover, the independence of phonon width on film thickness indicates consistent film quality across different thicknesses, ruling out any defects associated with carrier trapping processes.[23] Furthermore, the effective e-ph coupling strength is extracted from the integrated intensity, following the method described in Reference [27]. Figure 1f illustrates the effective e-ph coupling strength as a function of film thickness. It is observed that the coupling strength increases and tends to saturate as the thickness of the film increases.

## 2.2. Interplay between Thickness and Photoexcitation

To shed light on the diminished e-ph coupling strength in the thinner films, we then conducted degenerate pump-probe measurements at low pump fluence levels (0.61 μJ/ cm$^2$ ~ 12.13 μJ/cm$^2$) using an Oscillator (Maitai SP) with 84 MHz repetition rate and 800 nm central wavelength. Both pump and probe beams were focused onto the same spot of the sample by an objective lens. The pulse width of both pump and probe pulses was around 150 fs at the sample surface. These pump fluences correspond to photoexcited carrier densities of 0.65 × 10$^{18}$ cm$^{-3}$ to 12.86 × 10$^{18}$ cm$^{-3}$ (More details can be found in Supporting Information.), which were calculated using the transfer matrix method.[28-30] The calculated light penetration depth is 82 nm~154 nm, which is much greater than the thickness of thin films. The detailed calculation process can be found in Supporting Information. Figure 2a displays the transient reflectivity change (Δ$R$/$R_0$) observed in 4.62 nm-thick Bi$_2$O$_2$Se thin film at various pump fluences. The signals initially decrease rapidly and then gradually return to the baseline, indicating photoexcited ground-state bleaching followed by a relaxation process. The negative peak values from Figure 2a as a function of pump fluences are plotted in Figure 2b with the corresponding linear fitting, indicating linear excitation under these measurement conditions.

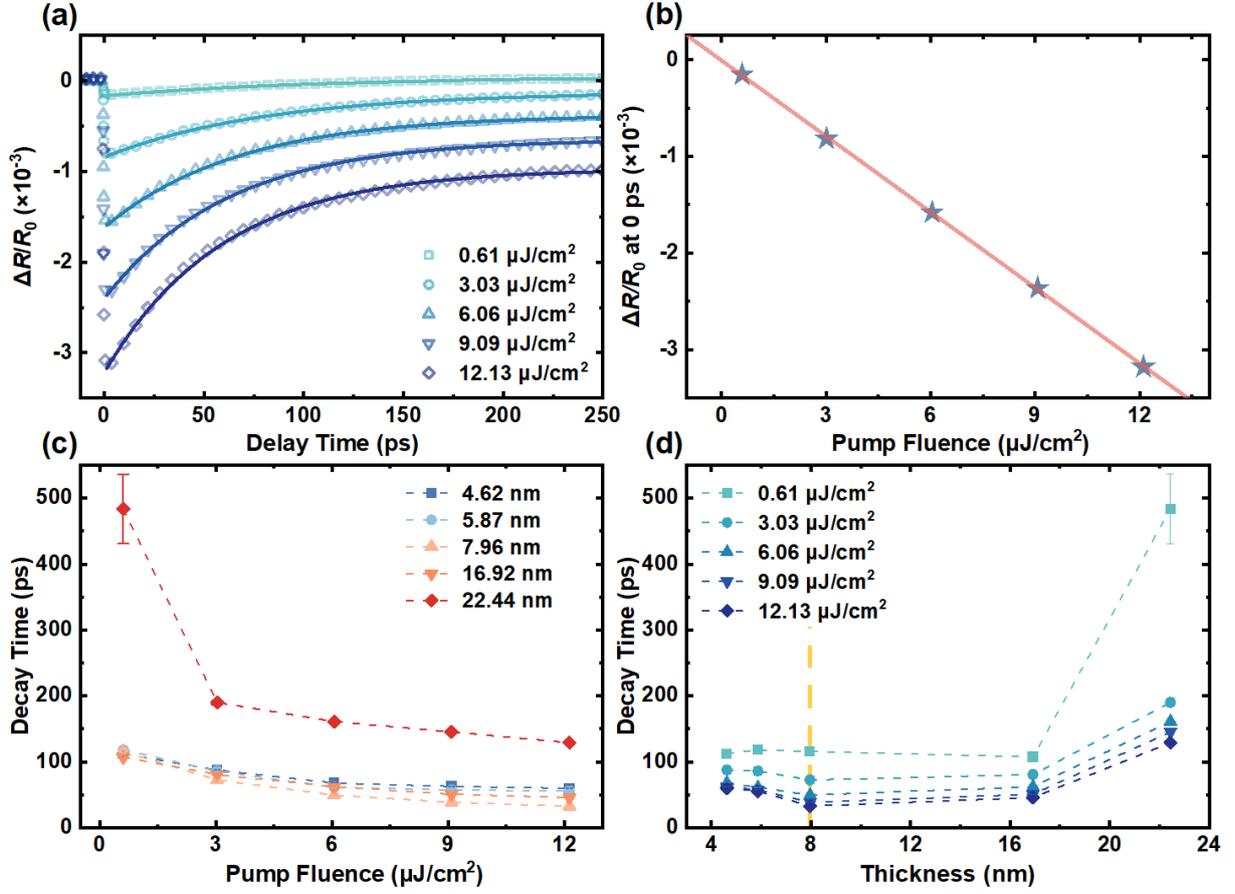

**Figure 2**. Ultrafast carrier dynamics at low pump fluences. a) Transient reflectivity change at pump fluences ranging from 0.61 μJ/cm² to 12.13 μJ/cm², along with their corresponding exponential fitting (solid lines) for 4.62 nm-thick $Bi_2O_2Se$ thin film. b) The negative peak values of $\Delta R/R_0$ (blue star) from Figure 2a as a function of pump fluences, and a linear fit (red line) is included. c) The extracted decay time as a function of pump fluence at various film thicknesses. d) The extracted decay time as a function of film thickness at various pump fluences. The yellow dashed line is for eye guidance at 7.96 nm.

In order to analyze the relaxation process observed in Figure 2a, we applied a single exponential decay fitting function as shown in Equation 1 (n=1).

$$\frac{\Delta R}{R_0} = \sum A_n e^{-t/\tau_n} + y_0 \qquad 1$$

where $\Delta R$ represents the transient reflectivity change at time $t$, $R_0$ is reflectivity without pump, $A_n$ is the amplitude of the corresponding process, $\tau_n$ is the decay time of different decay channels, and $y_0$ is an offset term.

In the pump fluence range of 0.61 μJ/cm$^2$ to 12.13 μJ/cm$^2$, only a single exponential decay function (Equation 1, n=1) is required for successful fitting. This suggests that only one electron decay channel is active under the measurement conditions employed.[31] The extracted decay time as a function of pump fluence is plotted in Figure 2c. The decay time for all samples exhibits a decreasing trend as the pump fluence increases. This slow decay process could be attributed to electron-hole (e-h) nonradiative recombination, "excimer exciton" dissociation (which occurs at approximately 2.09 eV),[21] and e-ph scattering (which typically occurs on the scale of a few picoseconds).[32] Interestingly, we found that the behavior is not monotonic when comparing the decay time with different film thicknesses, as shown in Figure 2d. In the region where the film thickness exceeds 7.96 nm, the decay time decreases as the thickness decreases, indicating the increasing importance of interface scattering in thinner films.[6] However, in the region where the film thickness is below 7.96 nm, the trend is opposite, suggesting the dominance of a different mechanism in this region.

## 2.3. Manifestation of Ferroelectricity at High Pump Fluence

To unravel the abnormal mechanism in the thinner films, we conducted high-fluence degenerate pump-probe experiments with an Amplifier (Ascend, 800 nm, 35 fs, 5 kHz). The pump fluence varies from 0.96 mJ/cm$^2$ to 4.08 mJ/cm$^2$, corresponding to photoexcited carrier density of 0.25 × 10$^{21}$ cm$^{-3}$ to 1.07 × 10$^{21}$ cm$^{-3}$. As shown in Figure 3a, the high-fluence signals exhibit a different trend compared to the low-fluence signals observed in Figure 2a. In this case, a double-exponential decay fitting function is required instead of a single-exponential decay function. This suggests the presence of multiple decay channels or relaxation processes under high-fluence excitation. Furthermore, from the inset of Figure 3a, we can observe that the peak values tend to exhibit a sublinear behavior with increasing pump fluences. This non-linear relationship between the peak values and pump fluences may indicate the influence of non-linear optical processes or excitonic interactions in thin films at high carrier densities.

These findings suggest that the underlying mechanisms in the thinner films at high carrier densities are more complex and involve additional processes beyond what a single-exponential decay model could explain. Further investigations and analysis are necessary to fully understand these phenomena.

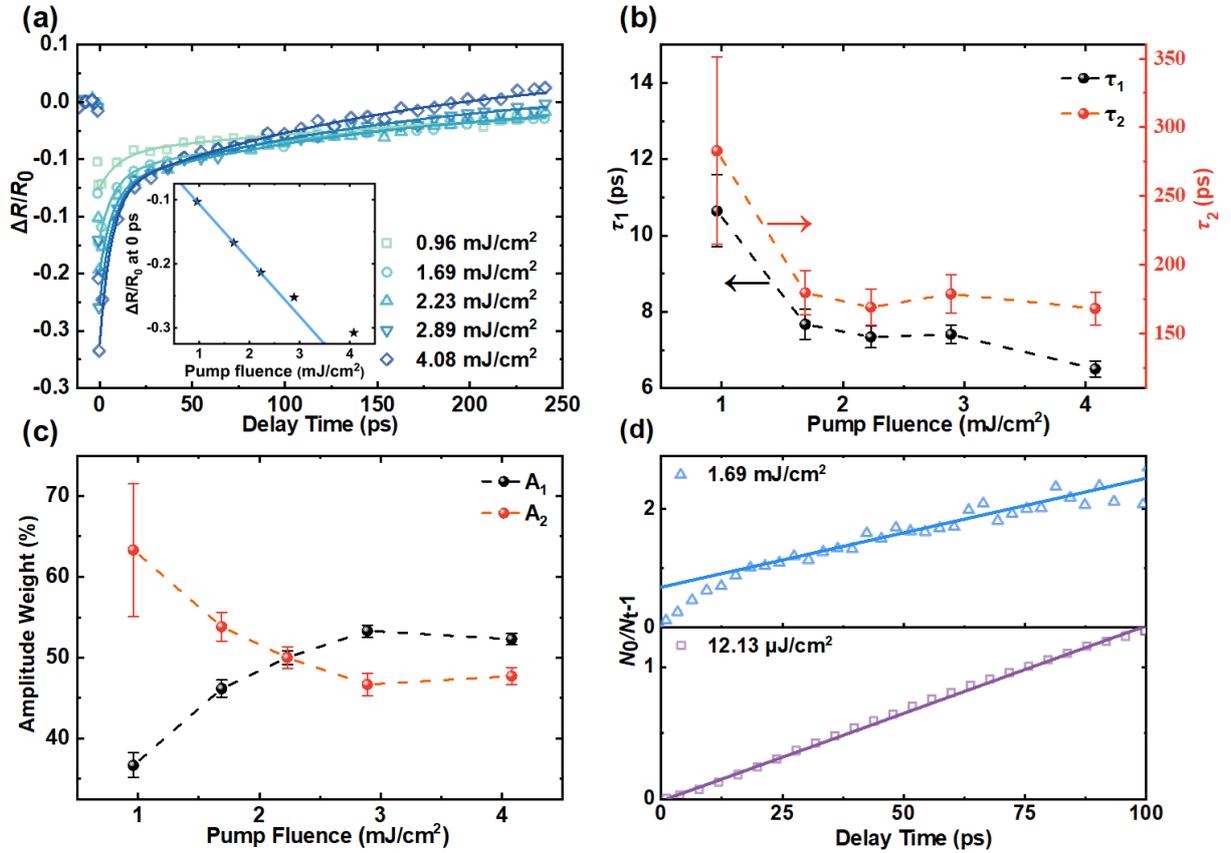

**Figure 3**. Ultrafast carrier dynamics at high pump fluences. a) Transient reflectivity change at pump fluences ranging from 0.96 mJ/cm$^2$ to 4.08 mJ/cm$^2$ for 4.62 nm-thick Bi$_2$O$_2$Se thin film, with their corresponding exponential decay fitting (solid lines). The inset displays the fluence-dependent peak values of $\Delta R/R_0$, and the blue line is an eye-guidance for a linear relation. b) The extracted decay time as a function of pump fluence using a double exponential decay fitting function. c) The corresponding amplitudes as a function of pump fluences. d) The time evolution of the quantity $N_0/N_t-1$ is shown, with empty purple squares representing the values obtained from Figure 2a (lower panel) and empty blue triangles representing the values obtained from Figure 3a (upper panel). The corresponding linear fits are represented by purple and orange lines, respectively.

By fitting the high-fluence data with a double exponential function (Equation 1, n=2), we can extract two decay time constants, denoted as $\tau_1$ and $\tau_2$, where $\tau_1$ represents the fast decay component, and $\tau_2$ represents the slow decay component. As shown in Figure 3b, both decay times decrease as the pump fluence increases. The slow decay process remains on the same order of magnitude as observed at low pump fluences, while the fast decay process ($\tau_1$) occurs within several picoseconds.

The fast decay process ($\tau_1$) can be attributed to several possible mechanisms: i) Carrier-carrier scattering, ii) Carrier-phonon scattering, and iii) Auger recombination at high pump intensities. However, it is important to note that carrier-carrier scattering typically happens on the timescale of hundreds of femtoseconds.[32, 33] Moreover, previous studies by Liu et al. demonstrated that the decay processes in Bi$_2$O$_2$Se are independent of temperature within the range of 80-300K,[22] which then cannot be due to carrier-phonon scattering. Furthermore, the fast decay process becomes more prominent at higher pump fluences, as depicted in Figure 3c, which is a signature of Auger recombination that is strongly linked with the carrier density. Therefore, we can conclude that Auger recombination dominates this fast decay process ($\tau_1$). These findings provide valuable insights into the underlying dynamics of the thin films under intense excitation conditions.

To further understand the decaying mechanisms for $\tau_1$ and $\tau_2$, we analyzed the rate equation (Equation 2) for Bi$_2$O$_2$Se, as shown in Figure 3d.[34]

$$\frac{dN_t}{dt} = C_1 N_t + C_2 N_t^2 + C_3 N_t^3 \qquad 2$$

where $N_t$ is the photogenerated carrier density, and the coefficients $C_1$, $C_2$, and $C_3$ depict the weight or rate constants for each process. This equation describes the time evolution of the photogenerated carrier density ($N_t$) following photoexcitation of the thin film. The first term represents first-order Shockley-Reed recombination, which is mediated by trap states. The second term corresponds to second-order non-geminate/free carrier recombination, while the third term represents three-body Auger recombination.[35-37]

It is well known that if carrier dynamics follow a bimolecular process, $N_0/N_t - 1$ will be linearly proportional to t, whereas $(N_0/N_t)^2 - 1$ will be linearly proportional to t for three-body Auger recombination,[38] where $N_0$ represents the initial carrier density. As shown in Figure 3d, the 4.62 nm sample exhibits non-linear behavior within 15 ps at a pump fluence of 1.69 mJ/cm$^2$, but becomes linear beyond that time (> ~15 ps). However, it consistently maintains linear behavior at a lower pump fluence of 12.13 μJ/cm$^2$ through the whole timescale. This further confirms that the fast decay channel ($\tau_1$) corresponds to three-body Auger recombination, while the slow decay channel ($\tau_2$) corresponds to e-h recombination. At the same time, this also confirms that at low pump fluences, the decay time is attributed to the recombination of e-h pairs. The relationship between $(N_0/N_t)^2 - 1$ and time t can be observed in Figure S4 (Supporting Information).

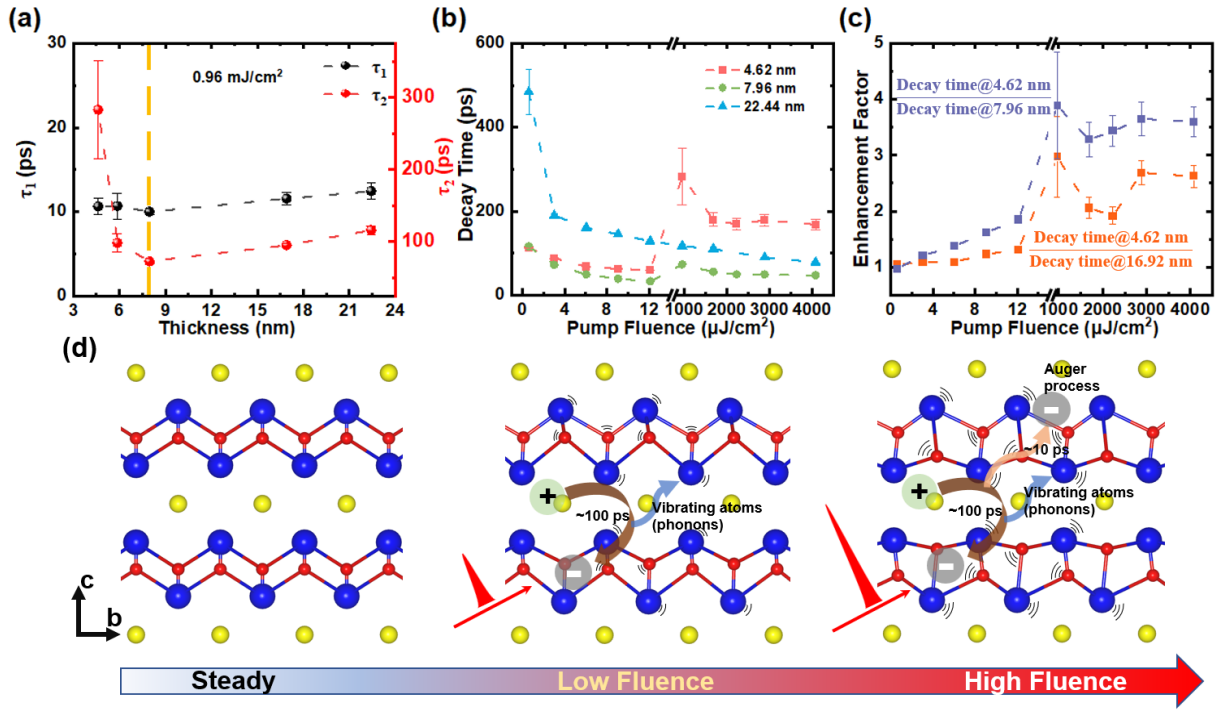

**Figure 4**. Schematics of carrier decaying channels. a) $\tau_1$ and $\tau_2$ as a function of film thickness at a pump fluence of 0.96 mJ/cm$^2$. The yellow dashed line is provided as a reference at 7.96 nm. b) The slow decay time as a function of pump fluences for three different film thicknesses: 4.62 nm (pink square), 7.96 nm (green circle), and 22.44 nm (blue triangle). c) The ratio of decay time@4.62 nm to decay time@7.96 nm and decay time@4.62 nm to decay time@16.92 nm, represented by purple square and orange square, respectively. d) A schematic illustration of light-induced ferroelectric transition mechanisms and dynamics in Bi$_2$O$_2$Se films, where "-" denotes the electron and "+" denotes the hole. Bi, O, and Se atoms are represented by blue, red, and yellow spheres, respectively.

To better understand the anomalous trends in the thinner films, as shown in Figure 2d, we plotted the thickness-dependent decay time at 0.96 mJ/cm$^2$ (Figure 4a) compared to Figure 2d. We observe that $\tau_1$ is relatively independent of film thickness, consistent with the argument that the Auger recombination process only depends on carrier density. On the other hand, $\tau_2$ exhibits a similar trend to Figure 2d, but with a more pronounced pattern. This substantial difference strongly suggests that the photoexcited carriers in the thinner films activate or manifest a unique mechanism. In thinner films, the exciton binding energy is known to be larger, leading to a shorter carrier lifetime.[26, 39, 40] However, this contradicts our observations in the thinner films. Therefore, we can eliminate the influence of thickness-dependent exciton binding energy on the observed phenomenon. The distinct behavior observed for $\tau_2$ in these thinner films indicates the presence of specific processes or interactions that are either absent or less prominent in thicker films.

To further understand this phenomenon, we have plotted $\tau_2$ against pump fluence for three different film thicknesses of 4.62 nm, 7.96 nm, and 22.44 nm, which is shown in Figure 4b. In the thicker films, $\tau_2$ decreases monotonically with increasing pump fluence, while in the thinner film, $\tau_2$ initially decreases at the low pump fluences, then abruptly increases at the high pump fluences, followed by a subsequent decrease.

Three possible mechanisms leading to prolonged decay time in the thinner films are i) dielectric screening, ii) Mott-like phase transition, and iii) a ferroelectric phase transition of $Bi_2O_2Se$ induced by symmetry breaking.[14] Regarding dielectric screening, there are two main sources to consider. The first is interface screening between the material and the substrate.[41] However, this effect does not vary with pump fluence. The second source is carrier-carrier screening.[42] As shown in Figure 4c, we found that the enhancement factor of the e-h recombination time between 4.62 nm and 7.96 nm, and we found this factor is approximately 3.89 at high fluences, despite the carrier density in the 4.62 nm film being only about 1.72 times greater than that in the 7.96 nm film. Furthermore, as the pump fluence increases, we noted a decrease in the e-h recombination time as shown in Figure 4b, which contradicted the carrier-carrier screening hypothesis. Therefore, we ruled out the dielectric screening. Regarding the Mott-like phase transition, the carrier lifetime is expected to be elongated,[43] however, the reported critical concentration is well below the photoexcited carrier density utilized in our experiments.[44] Hence, we attribute the prolongation of decay time in the thinner $Bi_2O_2Se$ films to a ferroelectric phase transition occurring in the thinner photo-excited active layer.

As illustrated in Figure 4c, at low pump fluences, the enhancement factor of the e-h recombination time between 4.62 nm and 16.92 nm slightly exceeds 1. According to the suggestion by Zhu et al., this enhanced e-h recombination time observed in the thinner films can be attributed to a ferroelectric transition induced by displacive symmetry breaking, leading to higher carrier mobility.[19] The compressive stress exerted by mica on $Bi_2O_2Se$ might be responsible for triggering the appearance of a ferroelectric phase in the thinner film. Additionally, under high pump fluences, with more photoexcited carriers reaching a non-linear excitation level, as depicted in Figure 4d, a greater lattice distortion occurs in the crystal structure, further promoting the ferroelectric phase transition. Therefore, at high pump fluences, the lifetime significantly increases for films with thicknesses less than 7.96 nm. The detailed dynamics are depicted in Figure 4d. The three diagrams in Figure 4d illustrate the atomic structure at steady state, low fluence excitation state, and high fluence excitation state, respectively. When subjected to low pump fluences, the electron experiences only one decay channel, e-h recombination. However, when exposed to high pump fluences, the electron exhibits two decay channels: (i) fast decay through Auger recombination and (ii) slow decay through e-h recombination.

Light-driven temporal phase transitions have been reported in bismuth by modifying the potential energy surface [30, 45] and 2D perovskites.[46] Furthermore, ultrafast spectroscopy has proven to be a useful tool for revealing the hidden meta-phase inaccessible in equilibrium phase diagrams.[33] In this study, we demonstrate that light-induced ferroelectric phase transitions occur more readily in thinner films, and this transition becomes apparent at higher photoexcited carrier densities.

## 3. Conclusion

Overall, we have investigated the thickness-dependent Raman spectroscopy and ultrafast carrier dynamics in the CVD-grown $Bi_2O_2Se$ thin films on mica substrate with thicknesses ranging from 22.44 nm to 4.62 nm at low and high pump fluence regions. We identified two distinct decay processes: a fast decay associated with Auger recombination and a slow decay corresponding to e-h recombination. The different trends of the slow decay time with pump fluence in the thinner and thick films suggest a unique mechanism affecting e-h recombination in the thinner films. By combining the thickness-dependence and fluence-dependence of the slow decay time, we demonstrate a ferroelectric transition in the thinner $Bi_2O_2Se$ films (< 8 nm) under the compressive strain from the substrate, while this transition can be manifested under highly non-equilibrium states. These findings deepen our understanding of the interplay between the ferroelectric phase and semiconducting properties of $Bi_2O_2Se$ thin films, providing a new route to manipulate the ferroelectric transition.

## 4. Experimental Section/Methods

*CVD Synthesis of $Bi_2O_2Se$ Nanoplates on Mica:* The controllable synthesis of ultra-thin to a few nanometers thick $Bi_2O_2Se$ flakes was effectuated following our previously reported vapor-solid growth technique.[24] Briefly, 2D $Bi_2O_2Se$ flakes were achieved on a mica substrate downstream 7–10 cm from the hot center in a 2 in. quartz tube. The source powder in the middle of the tube furnace was heated at 700 °C to transfer $Bi_2O_2Se$ vapors. The timeframe for the successful growth of $Bi_2O_2Se$ flakes lasted 10 minutes, and ultra-pure Ar gas was deployed with a flow rate of 220 sccm.

*Optical Characterizations:* The thickness of $Bi_2O_2Se$ nanosheets was measured by atomic force microscopy (AFM Bruker Dimension Icon, Germany). Raman results were collected on a Renishaw InVia Raman Microscope system with a 532 nm excitation laser in backscattering geometry.

*Degenerate pump-probe system*: In this work, we utilized two home-built degenerate femtosecond laser pump-probe systems for transient measurements of carrier dynamics under ambient conditions. The time delay was controlled using a mechanical delay stage. At the low pump fluences, 800 nm laser pulses with a pulse width of 35 fs and a repetition rate of 84 MHz from the oscillator (Maitai SP, Spectra Physics) were split into pump pulses and probe pulses using a beam splitter. The spot diameter of the pump beam is approximately 50 μm, while the spot diameter of the probe beam is approximately 18 μm ($1/e^2$). The pump pulses were modulated at a frequency of 2 kHz using a mechanical chopper, and the signal from the probe pulses was recorded using a detector (Thorlabs, PDA100) linking to a lock-in amplifier (Stanford Research Systems, SR860). The absorption depth of 800 nm in $Bi_2O_2Se$ is much larger than the sample thicknesses we studied. At the high pump fluence, the light source was switched to an Amplifier (Ascend, 800 nm, 35 fs, 5 kHz). The spot diameter of the pump beam is approximately 13 μm, while the spot diameter of the probe beam is approximately 10 μm ($1/e^2$). The pump pulses were modulated at a frequency of 585 Hz using a mechanical chopper, and the probe signal was recorded using a balanced photodiode (Thorlabs, PDB210A/M) linking to a lock-in amplifier (Stanford Research Systems, SR860).

Supporting Information is available from the Wiley Online Library or the author.

**Acknowledgments**
F. H. acknowledges fruitful discussions with Prof. Xiaolong Zou. The authors acknowledge supports from the "National Natural Science Foundation of China" (Grant Nos. 62204071, U22A2093, 22250410269, 52350610261), the Guangdong Basic and Applied Basic Research Foundation (Grant Nos. 2021A1515110606, 2023A1515011387), "China Postdoctoral Science

Foundation" (Grant No. 11110032542301) and Shenzhen Science and Technology Program (RCBS20210609103822042).

# Electronic Supplementary Material

## Hidden phase uncovered by ultrafast carrier dynamics in thin $Bi_2O_2Se$

Hao Li,[a,b] Adeela Nairan,[c] Xiaoran Niu,[a,b] Yuxiang Chen,[d] Huarui Sun,[d] Linqing Lai,[e] Jingkai Qin,[e] Leyang Dang,[f] Guigen Wang,[f] Usman Khan*[c] and Feng He*[a,b]

a. State Key Laboratory on Tunable Laser Technology, Harbin Institute of Technology (Shenzhen), Shenzhen 518055, P. R. China

b. Guangdong Provincial Key Laboratory of Aerospace Communication and Networking Technology, Harbin Institute of Technology (Shenzhen), Shenzhen 518055, P. R. China. E-mail: hefeng2020@hit.edu.cn

c. Institute of Functional Porous Materials, School of Materials Science and Engineering, Zhejiang Sci-Tech University, Hangzhou 310018, P. R. China. E-mail: usmankhan.2017@tsinghua.org.cn

d. School of Science and Ministry of Industry and Information Technology, Key Laboratory of Micro-Nano Opto-electronic Information System, Harbin Institute of Technology (Shenzhen), Shenzhen, Guangdong 518055, P. R. China

e. Sauvage Laboratory for Smart Materials, School of Materials Science and Engineering, Harbin Institute of Technology (Shenzhen), Shenzhen 518055, P. R. China

f. Shenzhen Key Laboratory for Advanced Materials, School of Materials Science and Engineering, Harbin Institute of Technology (Shenzhen), Shenzhen 518055, P. R. China

**S1. AFM images of Bi$_2$O$_2$Se thin films with different thicknesses.**

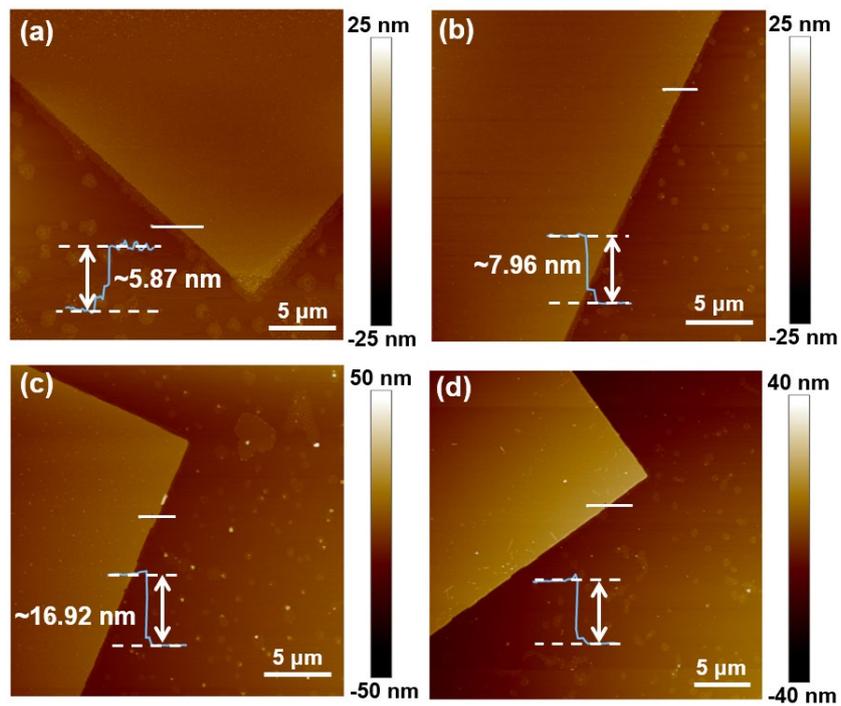

**Figure S1.** Additional AFM analysis of ultrathin Bi$_2$O$_2$Se films. AFM image of Bi$_2$O$_2$Se thin film with a thickness of (a) 5.87 nm, (b) 7.96 nm, (c) 16.92 nm, and (d) 22.44 nm.

**S2. Transfer matrix method calculation**

The transfer matrix method is commonly employed to determine the complex refractive index of materials.

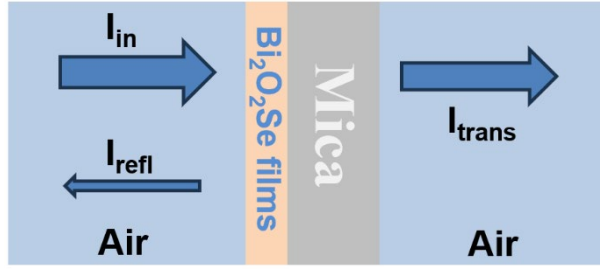

**Figure S2.** A diagram illustrating transfer matrix method.

Specifically, for the Air/Bi$_2$O$_2$Se and Bi$_2$O$_2$Se/Mica interfaces under consideration, the transfer matrices can be expressed as follows:

$$M_{interface} = \frac{1}{t}\begin{pmatrix} 1 & r \\ r & 1 \end{pmatrix} \qquad \text{S1}$$

$$t = \frac{2\bar{n_1}}{\bar{n_1}+\bar{n_2}} \qquad \text{S2}$$

$$r = \frac{\bar{n_1}-\bar{n_2}}{\bar{n_1}+\bar{n_2}} \qquad \text{S3}$$

Where $t$ and $r$ are the transmission and the reflection of electric field, respectively.

Where $\bar{n_1}$ and $\bar{n_2}$ are the complex refractive index of the materials on the front and back sides of the interface. Since the thickness of mica is much thicker than the sample thickness, we consider it to be semi-infinite. The laser propagation inside the Bi$_2$O$_2$Se is modeled by a propagation matrix as:

$$M_{propagation} = \begin{pmatrix} e^{-\frac{i2\pi \bar{n}d}{\lambda}} & 0 \\ 0 & e^{\frac{i2\pi \bar{n}d}{\lambda}} \end{pmatrix} \qquad \text{S4}$$

where $\bar{n}$ is the complex refractive index of Bi$_2$O$_2$Se, $d$ is the thickness, and $\lambda$ is the wavelength. So, the total transfer matrix can be written as:

$$M = M_{Air/Bi_2O_2Se} \times M_{propagation} \times M_{Bi_2O_2Se/Mica} = \begin{pmatrix} M_{11} & M_{12} \\ M_{21} & M_{22} \end{pmatrix} \qquad \text{S5}$$

The reflectance can be written as:

$$R = \left|\frac{M_{21}}{M_{11}}\right|^2 \qquad \text{S6}$$

The transmittance can be written as:

$$T = \left|\frac{1}{M_{11}}\right|^2 \times \frac{|n_{Mica}|}{|n_{Air}|} \qquad \text{S7}$$

With R and T from the experiment (Figure S2), the complex refractive index $\bar{n} = n + ik$ can be solved at each irradiance, where $n$ and $k$ are the real and imaginary part, respectively. Then, the absorption coefficient can be calculated as $\alpha = 4\pi k/\lambda$. Results are summarized in Table S1.

Next, the initial carrier density $N_0$ is given by $N_0 = (1-R-T) \times F / (d_{eff} \times E_{ph})$, where $F$ represents the pump fluence, $E_{ph}$ represents the photon energy. $d_{eff}$ stands for the effective absorption depth, which is equal to the sample thickness since the laser penetration depth is much greater than the thickness of the sample.[1]

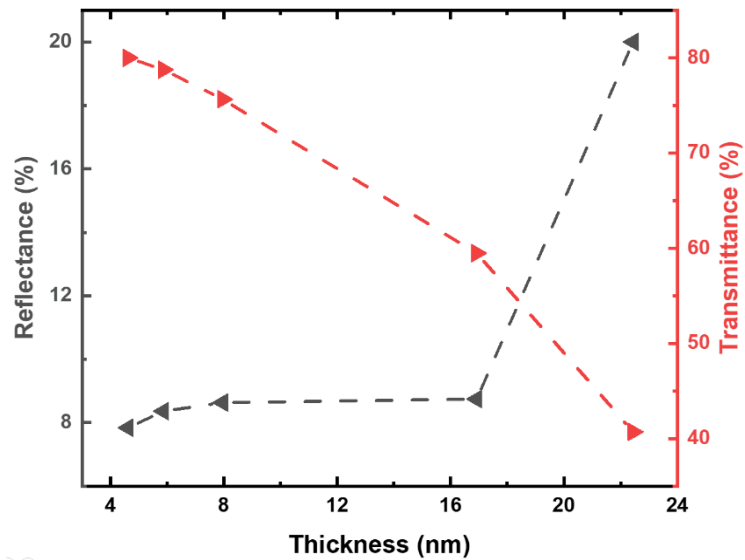

**Figure S3.** The experimentally measured transmittance and reflectance.

**Table S1.** The calculated values for the real and imaginary parts of the refractive index, as well as the absorption coefficient, are as follows.

| Thickness (nm) | Layer | n | k | $\alpha*10^4$ (cm$^{-1}$) |
| --- | --- | --- | --- | --- |
| 22.44 | 34 | 3.434 | 0.778 | 12.21 |
| 16.92 | 27 | 2.667 | 0.717 | 11.26 |
| 7.96 | 13 | 2.929 | 0.556 | 8.73 |
| 5.87 | 9 | 2.958 | 0.543 | 8.52 |
| 4.62 | 7 | 3.343 | 0.414 | 6.50 |

## S3. $N_0^2/N_t^2-1$ vs. t

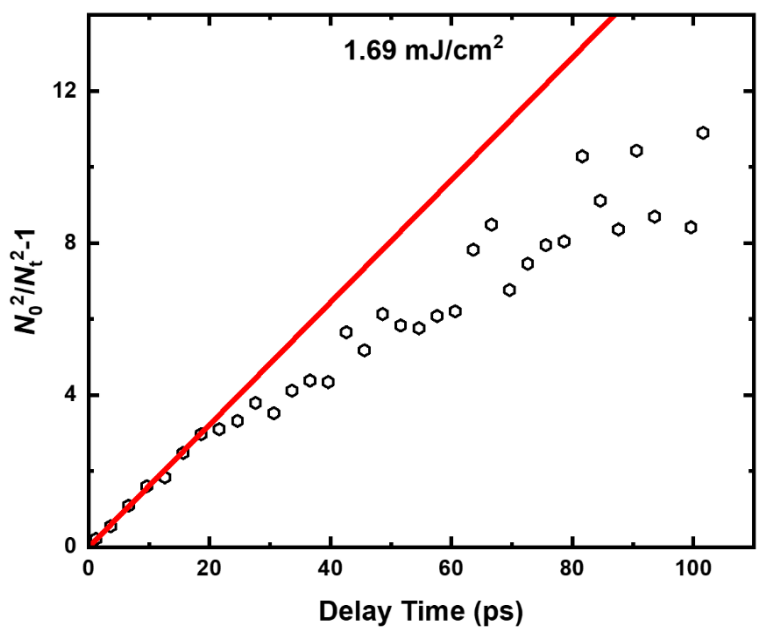

**Figure S4.** Plot of the relationship between $N_0^2/N_t^2-1$ and t at a thickness of 4.62 nm and a pump fluence of 1.69 mJ/cm². The red line represents the linear fit within the first 15 ps.

At a pump fluence of 1.69 mJ/cm², the 4.62 nm sample exhibits linear behavior for the first 15 ps, but becomes nonlinear after that time (> ~15 ps). This indicates that Auger recombination dominates at the first 15 ps.

## S4. Sample uniformity

We checked the AFM data and obtained the height fluctuation of the surface, as shown in Figure S5(a). The fluctuation is no more than 5 Å. In addition, when we carried out the pump-probe experiment, three different test points were selected on every sample to check the repeatability, as shown in Figure S5(b). The results show that the electronic signals of the three sample points have good repeatability, and fitted decay time of the three spots are 129.56 ± 0.75 ps, 129.38 ± 0.74 ps, and 130.22 ± 0.87 ps, respectively, agreeing well within error bar range.

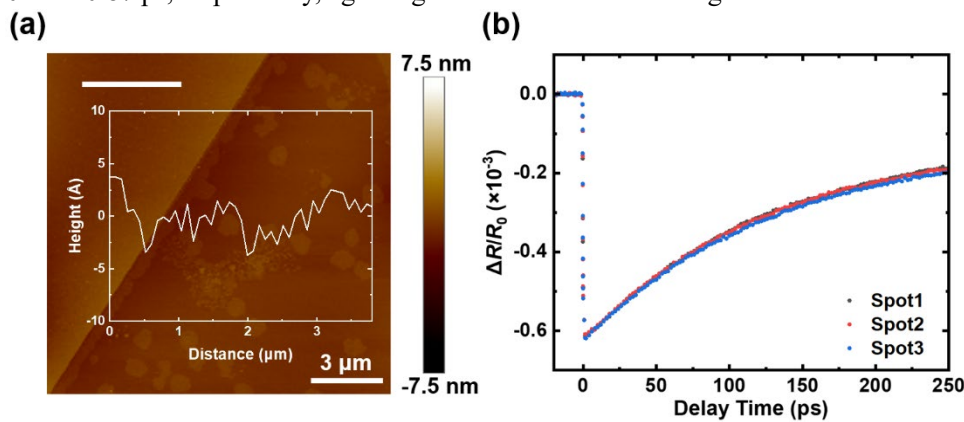

**Figure.S5** (a) The roughness of the sample surface. (b) Transient reflectance curves of three sample points for 22.44 nm sample at a pump fluence of 3.03 µJ/cm². The black, red and blue dots represent the replacement of three different sample points for samples of the same thickness.